\newcommand{\beq}{\begin{equation}}
\newcommand{\eeq}{\end{equation}}

\newcommand{\id}
 {i\kern.06em\hbox{\raise.25ex\hbox{$/$}\kern-.60em$\partial$}}

\newcommand{\bs}{/\kern-.52em b}
\newcommand{\qs}{/\kern-.52em s}

\newcommand{\cL}{{\cal L}}

\newcommand{\p}{\partial}
\newcommand{\yp}{^{\prime}}
\newcommand{\ts}{\tilde{e}}
\newcommand{\tp}{\tilde{\pi}}
\newcommand{\dd}
{\kern.06em\hbox{\raise.25ex\hbox{$/$}\kern-.60em$\partial$}}

\newcommand{\ep}{\epsilon}
\newcommand{\tr}{\mathop{\rm tr}\nolimits}

\documentstyle[12pt]{article}
\date{}
\begin{document}
\title{Superalgebra and Conservative Quantities in $N=1$ Complex
Supergravity\footnotetext{\# Corresponding
adress}
\thanks{On leave of absence from Physics Department, Shanghai
University, 201800, Shanghai, China}}
\author{{Sze-Shiang Feng$^{1,2,3,\#}$, Bin Wang$^4$}\\
1.{\small {\it High Energy Section, ICTP, Trieste,34100,Italy}}\\
e-mail:fengss@ictp.trieste.it\\
2.{\small {\it CCAST(World Lab.), P.O. Box 8730, Beijing 100080}}\\
3.{\small {\it Modern Physics Department, University of Science 
and Technology of China, 230026,
Hefei,China}}$^\#$\\e-mail:zhdp@ustc.edu.cn$^\#$\\
4.{\small {\it Physics Department, Shanghai Normal University,
100234, Shanghai, China}}}
\maketitle
\newfont{\Bbb}{msbm10 scaled\magstephalf}
\newfont{\frak}{eufm10 scaled\magstephalf}
\newfont{\sfr}{eufm7 scaled\magstephalf}
\input amssym.def
\baselineskip 0.3in
\begin{center}
\begin{minipage}{135mm}
\vskip 0.3in
\baselineskip 0.3in
\begin{center}{\bf Abstract}\end{center}
  {The $N=1$ self-dual supergravity has $SL(2,\Bbb C)$ symmetry
  and the
  left-handed and right-handed local supersymmetries. These
  symmetries
  result in $SU(2)$ charges as the angular-momentum and the
  supercharges. The model possesses also the invariance under
  the general translation transforms and this invariance
  leads to the energy-momentum. All the definitions
  are generally
  covariant. As the $SU(2)$ charges and the energy-momentum
  we obtained previously constituting the 3-Poincare
  algebra in
  the Ashtekar'complex gravity, the $SU(2)$ charges, the
  supercharges and the energy-momentum in simple
  supergravity also restore
  the super-Poincare algebra, and this serves to support
  the reasonableness of their interpretations.
  \\PACS number(s): 11.30.Cp,11.30.Pb,04.20.Me,04.65.+e.
   \\Key words: superalgebra, conservative quantities,
   supergravity}
\end{minipage}
\end{center}
\vskip 1in
\begin{center}
\section{Introduction}
\end{center}
\indent The study of self-dual gravities has drawn much
attention in the past decade since the discovery of Ashtekar's
new variables, in terms of which the constraints can be greatly
simplified\cite{s1}-\cite{s2}. The new phase variables
consist of  densitized $SU(2)$ soldering forms $\tilde{e}
^i\,_A\,^B$ from which a metric density is obtained according
to the definition $q_{ij}=-{\rm Tr}\tilde{e}_i\tilde{e}_j$,
and a complexified connection $A_{iA}\,^B$ which
carries the momentum
dependence in its imaginary part. The original
Ashtekar's self-dual
canonical gravity permits also a Lagrangian
formulation\cite{s3}
-\cite{s4}. The supersymmetric extension of this
Lagrangian
formulation, which is equivalent to the simple real
supergravity,
was proposed by Jacobson\cite{s5}, and the corresponding
Ashtekar complex canonical transform was given by
Gorobey et al\cite{s6}.\\
\indent  In our previous works, we have obtained
the $SU(2)$ charges
and the energy-momentum in the Ashtekar's formulation
of Einstein
gravity\cite{s7}-\cite{s8} and they are closely
related to the
angular-momentum\cite{s9}-\cite{s11} and the
energy-momentum \cite{s12}
in the vierbein formalism of Einstein gravity.
The fact that
the algebra formed by their Poisson brackets
{\it do}
constitute the 3-Poincare algebra on the Cauchy
surface
supports from another aspect that their
definitions are reasonable.\\
\indent Out of the same reason,
the definitions of $SU(2)$ charges, which
are to be interpreted as the angular-momentum,
the supercharges
and the energy-momentum are also interesting
and important
aspects in the simple self-dual supergravity.
In this paper,
we will exploit the $SL(2,\Bbb C)$ invariance,
the left-handed
and right-handed supersymmetry and the invarinace
under the
general translation transform\cite{s12} to
obtain the conservative
charges under consideration. This paper
is arranged as follows.
In section 2, we will give a brief review of
the $N=1$ self-dual
supergravity. In section 3, we will derive the
$SU(2)$ charges from
the original Lagrangian of Jacobson. In
section 4, we derive the energy-momentum from a
slightly different
Lagrangian and the general translation. In section 5,
we derive the
supercharges from the invariance under left-handed
and right-handed
local supersymmetric transforms. The last section
is devoted to
summary and discussions.
\begin{center}
\section{A Brief Review of the Model}
\end{center}
   The Lagrangian density is\cite{s5}
\beq
{\cal L}_J=\frac{1}{\sqrt{2}}(e^{AA^{\prime}}
\wedge e_{BA^{\prime}}\wedge F_A\,^B+
ie^{AA^{\prime}}\wedge\bar{\psi}_{A^{\prime}}
\wedge{\cal D}\psi_A)
\eeq
The dynamical variables are the real tetrad
$e^{AA^{\prime}}$ (the
"real" means $\bar{e}^{A\yp A}=e^{AA\yp}$), the
traceless left-handed $SL(2.\Bbb C)$ connection
$A_{\mu MN}$
and the complex anticommuting spin-$\frac{3}{2}$
gravitino
field $\psi_{\mu A}$. The $SL(2,\Bbb C)$ covariant
exterior
derivative is defined by
\beq
{\cal D}\psi_M:=d\psi_M+A_M\,^N\wedge\psi_N
\eeq
and the curvature 2-form is
\beq
F_M\,^N:=dA_M\,^N+A_M\,^P\wedge A_P\,^N
\eeq
The indices are lowered and raised with the antisymmetric
$SL(2, \Bbb C)$ spinor $\epsilon^{AB}$ and its
inverse $\epsilon_{AB}$
according to the convention $\lambda_B=
\lambda^A\epsilon_{AB},
\lambda^A=\epsilon^{AB}\lambda_B$, and the implied
summations are always
in north-westerly fashion: from  the left-upper
to the right-lower.
The Lagrangian eq.(1) is a holomorphic functions
of the
connection and the equation for $A_{\mu A}\,^B$
is equivalent
to
\beq
{\cal D}e^{AA^{\prime}}=\frac{i}{2}\psi^A
\wedge\bar{\psi}^{A^{\prime}}
\eeq
provided $e^{AA\yp}$ is real. The Lagrangian
$\frac{1}{2}({\cal L}
_J+\bar{\cal L}_J)$ for real supergravity is a
non-holomorphic function
but leads to no surfeit of field equations. Under
the left-handed
local supersymmetric transform generated by anticommuting
parametres $\ep_A$
\beq
\delta\psi_A=2{\cal D}\ep_A,\,\,\,\,\,\,
\delta\bar{\psi}_{A\yp}=0,\,\,\,\,\,\,\,\,
\delta e_{AA\yp}=-i\bar{\psi}_{A\yp}\ep_A
\eeq
the Lagrangian ${\cal L}_J$ is invariant {\it without}
using any one
of the Euler-Lagrangian equations while under the
right-handed transform
\beq
\delta\psi_A=0,\,\,\,\,\,\,\,\,\,
\delta\bar{\psi}_{A\yp}=2{\cal D}\bar{\ep}
_{A\yp},\,\,\,\,\,\,\,\,\,\, \delta e_{AA\yp}=
-i\psi_A\bar{\ep}_{A\yp}
\eeq
${\cal L}_J$ is invariant {\it modulo} the field equations.\\
\indent The (3+1) decomposition is effected as
\beq
{\cal L}_J=\ts^{kAB}\dot{A}_{kAB}+\tp^{kA}
\dot{\psi}_{kA}-{\cal H}
\eeq
\beq
{\cal H}:=e_{0AA\yp}{\cal H}^{AA\yp}+\psi_{0A}{\cal S}^A
+\hat{{\cal S}}^{A\yp}\bar{\psi}_{0A\yp}+A_{0AB}
{\cal J}^{AB}
+({\rm total\,\, divergence})
\eeq
The canonical momenta are
\beq
\ts^{kAB}:=-\frac{1}{\sqrt{2}}\ep^{ijk}e_i\,^{AA\yp}
e_j^B\,_{A\yp}
\eeq
\beq
\tp^{kA}:=\frac{i}{\sqrt{2}}\ep^{ijk}e_i\,^{AA\yp}
\bar{\psi}_{jA\yp}
\eeq
and the constraints are
\beq
{\cal H}^{AA\yp}:=\frac{1}{\sqrt{2}}\ep^{ijk}(e_i\,
^{BA\yp}F_{jkB}\,^A-i\bar{\psi}_i\,^{A\yp}
{\cal D}_j\psi_k\,^A)
\eeq
\beq
{\cal S}^A:={\cal D}_k\tp^{kA}
\eeq
\beq
\hat{\cal S}^{A\yp}:=\frac{i}{\sqrt{2}}
\ep^{ijk}e_i\,^{AA\yp}{\cal D}
_j\psi_{kA}
\eeq
\beq
{\cal J}^{AB}:={\cal D}_k\ts^{kAB}-\tp^{k(A}\psi_k\,^{B)}\
\eeq
\indent The 0-components $e_{0AA\yp}, \psi_{0A},
\bar{\psi}_{0A\yp}
$ and $A_{0AB}$ are just the Lagrange multipliers
and the dynamical
conjugate pairs are $(\ts^{kAB},A_{jAB}),
(\tp^{kA}, \psi_{kA})$.
The constraints ${\cal H}^{AA\yp}=0$ and
$\hat{\cal S}^{A\yp}=0$
generate the following two
\beq
\ddot{\cal H}^{AB}:=(\ts^j\ts^kF_{jk})^{AB}
+2\tp^j\ts^k{\cal D}_{[j}\psi_{k]}\ep^{AB}
+2(\tp^j{\cal D}_{[j}\psi_{k]})\ts^{kAB}=0
\eeq
\beq
{\cal S}^{\dag A}:=\frac{1}{\sqrt{2}}\ep^{ijk}
\ts_i\,^{AB}{\cal D}
_j\psi_{kB}=0
\eeq
The equations of motion will be properly expressed
in Hamiltonian form
$\dot{f}=\{H, f\} $if we assign the Poisson brackets
\beq
\{\ts^{kAB}(x),A_{jAB}(y\}=\delta_j\,^k\delta_{(M}
\,^A\delta_{N)}
\,^B\delta^3(x,y)
\eeq
\beq
\{\tp^{kA}(x), \psi_{jA}(y)\}=-\delta_j\,^k\delta_M\,
^A\delta^3(x,y)
\eeq
all other brackets among these quantities being zero.\\
\indent This is the outline of the theory.
\begin{center}
\section{The $SU(2)$ Charge}
\end{center}
Under any $SL(2,\Bbb C)$ transform
$$ e_{\mu AA\yp}\rightarrow L_A\,^B\bar{R}_{A\yp}\,^{B\yp}
e_{\mu BB\yp},\,\,\,\,\,\,\, \psi_A\rightarrow L_A\,^B\psi
_B,\,\,\,\,\,\,\, \bar{\psi}_{A\yp}\rightarrow
\bar{R}_{A\yp}\,^{B\yp}\bar{\psi}_{B\yp} $$
\beq
A_{\mu MN}\rightarrow L_M\,^AA_{\mu A}\,^{B}
(L^{-1})_{BN}+L_M\,^A\p_{\mu}(L^{-1})_{AN}
\eeq
${\cal L}_J$ is invariant. $L$ and $\bar{R}$ may not
neccessarily
related by complex conjugation. Note that
$L_{AB}=-(L^{-1})_{BA}$,
the transform of $A$ may also be written as
\beq
A_{\mu MN}\rightarrow L_M\,^A
L_N\,^BA_{\mu AB}-L_M\,^A\p_{\mu}L_{NA}
\eeq
For infinitesimal transform, $L_A\,^B=\delta_A\,^B+\xi_A\,^B$
where $\xi_{AB}=-\xi_{BA}$ are infinitesimal parametres. Thus
we have
\beq
\delta_{\xi}A=[\xi, A]-d\xi,\,\,\,\,\,\,\,\,
\delta \psi=\xi\psi
\eeq
When calculating the variation of the Lagrangian, one must take
into consideration of the anticommuting feature of the gravitino
field. We write the variation in the way that
\beq
\delta{\cal L}_J=\delta\phi^A(\frac{\p}{\p\phi^A}
-\p_{\mu}\frac{\p}{\p\p_{\mu}\phi^A}){\cal L}_J+\p_{\mu}
(\delta\phi^A\frac{\p}{\p\p_{\mu}\phi^A}{\cal L}_J)
\eeq
where $\phi^A$ denotes any field involved in the first order
Lagrangian. Now both $\frac{\p}{\p\phi^A}$ and
$\frac{\p}{\p\p_{\mu}\phi^A}$
are (anti-)commuting if $\phi^A$ is (anti-)commuting,
and so there is
no ordering problem.\\
\indent The invariance of ${\cal L}_J$ under the
infinitesimal $SL(2,\Bbb C)$ transform is equivalent to the
following modulo the field equations
\beq
\p_{\rho}(\delta A_{\sigma A}\,^B
\frac{\p{\cal L}_J}{\p\p_{\rho}A_{\sigma A}\,^B}
+\delta\psi_{\sigma A}\frac{\p{\cal L}_J}{\p\p_{\rho}
\psi_{\sigma A}})=0
\eeq
For constant $\xi$, we have
\beq
\p_{\rho}(\frac{1}{\sqrt{2}}\ep^{\mu\nu\rho\sigma}
e_{\mu}\,^{AA\yp}
e_{\nu BA\yp}[\xi, A_{\sigma}]_A\,^B
+\frac{i}{\sqrt{2}}\ep^{\mu\nu\rho\sigma}e_{\mu}\,^{AA\yp}
\bar{\psi}_{\nu A\yp}(\xi\psi_{\sigma})_A)=0
\eeq
we have therefore the conservation of $SU(2)$ charges
\beq
\p_{\mu}\tilde{j}^{\mu}_{AB}=0
\eeq
where
$$
\tilde{j}^{\rho}_{AB}=\frac{1}{\sqrt{2}}\ep^{\mu\nu\rho\sigma}
(e_{\mu A}\,^{A\yp}e_{\nu MA\yp}A_{\sigma B}\,^M
-e_{\mu}\,^{MA\yp}e_{\nu BA\yp}A_{\sigma MA}$$
\beq
+\frac{i}{2}e_{\mu A}\,^{A\yp}\bar{\psi}_{\nu A\yp}
\psi_{\sigma B}
+\frac{i}{2}e_{\mu B}\,^{A\yp}\bar{\psi}_{\nu A\yp}
\psi_{\sigma A})
\eeq
Thus
\beq
J_{AB}=\int_{\Sigma}\tilde{j}^0_{AB} d^3x
\eeq
where
$$
\tilde{j}^0_{AB}=\frac{1}{\sqrt{2}}\ep^{ijk}
(e_{i A}\,^{A\yp}e_{j MA\yp}A_{k B}\,^M
-e_{i}\,^{MA\yp}e_{j BA\yp}A_{k MA}$$
\beq
+\frac{i}{2}e_{i A}\,^{A\yp}\bar{\psi}_{j A\yp}\psi_{kB}
+\frac{i}{2}e_{i B}\,^{A\yp}\bar{\psi}_{j A\yp}\psi_{kA})
\eeq
Using eq(9) and eq(10), $\tilde{j}^0_{AB}$ can be written as
\beq
\tilde{j}^0_{AB}=[\tilde{e}^k, A_k]_{AB}+\tp_{k(A}\psi^k_{B)}
\eeq
The constraint ${\cal J}_{AB}=0$ guarantees that
\beq
J_{AB}\approx\int_{\Sigma}\p_k\tilde{e}^k_{AB}=\int_{\p\Sigma}
\tilde{e}^k_{AB}ds_i
\eeq
where $ds_i=\frac{1}{2}\ep_{ijk}dx^j\wedge dx^k$.
It can also be obtained in the following way. Using the field
equation $e^{A\yp(A}\wedge({\cal D}e^{B)}\,_{A\yp}
-\frac{i}{2}\psi^{B)}\wedge\bar{\psi}_{A\yp})=0$, we have
$$\ep^{\rho\mu\nu\sigma}[e_{\mu A}\,^{A\yp}(\p_{\sigma}
e_{\nu BA\yp}
+A_{\sigma B}\,^Me_{\nu MA\yp}+\frac{i}{2}
\bar{\psi}_{\nu A\yp}
\psi_{\sigma B})$$
\beq
+e_{\mu B}\,^{A\yp}(\p_{\sigma}e_{\nu AA\yp}
+A_{\sigma A}\,^Me_{\nu MA\yp}+\frac{i}{2}
\bar{\psi}_{\nu A\yp}
\psi_{\sigma A})]=0
\eeq
so
\beq
\tilde{j}^{\rho}_{AB}=-\frac{1}{\sqrt{2}}
\ep^{\rho\mu\nu\sigma}
\p_{\sigma}(e_{\mu A}\,^{A\yp}e_{\nu BA\yp})
\eeq
Using
\beq
e_{[\mu A}\,^{A\yp}e_{\nu]BA\yp}=e_{[\mu AC}e_{\nu]B}\,
^C-i\sqrt{2}
n_{[\mu}e_{\nu]AB}
\eeq
we have
\beq
\begin{array}{rcl}
\tilde{j}^0_{AB}&=&-\frac{1}{\sqrt{2}}\ep^{ijk}
\p_k(e_{[iA}\,^{A\yp}e_{j]BA\yp})
=-\frac{1}{\sqrt{2}}\ep^{ijk}\p_k
(e_{[iAC}e_{j]B}\,^C-i\sqrt{2}n_{[i}e_{j]AB})\\\\
&=&\frac{1}{\sqrt{2}}\ep^{ijk}\p_k(e_ie_j)_{AB}
=\p_k\ts^{k}_{AB}
\end{array}
\eeq
which is exactly the same as eq.(30)
We can thus
have the Poisson brackets
$$
\{J_{AB}, J_{MN}\}=\{\int_{\p\Sigma}\tilde{e}^k_{AB}ds_k,\,\,\,\,
\int_{\Sigma}(\tilde{e}^i\,_M\,^PA_{iPN}+
\tilde{e}^i\,_N\,^PA_{iPN})d^3x\}$$
\beq
=\frac{1}{2}(J_{MA}\ep_{NB}+J_{MB}\ep_{NA}+J_{NA}\ep_{MB}
+J_{MA}\ep_{NB})
\eeq
Now the flat dreibein on $\Sigma$ is needed
in order to find the angular-
momentum $J_i$. To clarify the notions, we use
the following conventions:
$\mu,\nu,...$ denote the 4-dim curved indices and
$i,j,k,$ denote
the 3-dim curved indices on $\Sigma$; $a,b,c,...$
denote the flat 4-dim
indices and $ l,m,n,...$ denote the flat 3-dim
indices on $\Sigma$.
The rigid flat vierbein is denoted as $E^a_{AA\yp}$
and the rigid flat
dreibein is denoted by $E^m_{AB}$. Then define
\beq
J_m:=\frac{1}{\sqrt{2}}E_m^{AB}J_{AB}
\eeq
and using the relation $\ep^{mnl}E_mE_n=\sqrt{2}E^l$ we have
\beq
\{J_m,J_n\}=\ep_{mnl}J^l
\eeq
 Therefore the $su(2)$ algebra is restored.
One may doubt the finiteness of $J_{MN}$ for isolated
systems. They are indeed finite because, in the
non-supersymmetric case, $J_{MN}$ is related to $J_{ab}$
by a linear transform\cite{s8}, where $J_{ab}$
is the angular-mometum obtained in the vierbein formalism
and are proved finite for general isolated systems, further,
it can give the correct formula of radiation of angular-momentum.
\cite{s9}-\cite{s10}.
As in the non-supersymmetric case\cite{s8}, we can also
obtain only
the $SU(2)$ charges instead of the whole $SL(2,\Bbb C)$
charges.
Yet, the angualr-momentum $J_{ab}$ obtained in\cite{s9}-
\cite{s10} is
completely contained in $J_{MN}$ (see section 5).
\begin{center}
\section{The Energy-momentum}
\end{center}
In order to obtain the energy-momentum, we do not
exploit the Lagrangian
${\cal L}_J$. Instead,we use the following ${\cal L}$
$$
{\cal L}=\frac{1}{\sqrt{2}}\ep^{\mu\nu\rho\sigma}
[-\p_{\rho}(e^{AA\yp}_{\mu}e_{\nu BA\yp})
A_{\sigma A}\,^B+e^{AA\yp}_{\mu}e_{\nu BA\yp}A_{\rho A}\,^M
A_{\sigma M}\,^B +\frac{i}{2}e^{AA\yp}_{\mu}
\bar{\psi}_{\nu A\yp}{\cal D}
_{\rho}\psi_{\sigma A}$$
\beq
\frac{i}{2}(\p_{\rho}e^{AA\yp}_{\mu}\bar{\psi}_{\nu A\yp}
\psi_{\sigma A}
+e^{AA\yp}_{\mu}\p_{\rho}\bar{\psi}_{\nu A\yp}\psi_{\sigma A})
+\frac{i}{2}e^{AA\yp}_{\mu}\bar{\psi}_{\nu A\yp}
\psi_{\sigma B}A_{\rho A}\,^B]
\eeq
i.e.
\beq
{\cal L}={\cal L}_J-\frac{1}{\sqrt{2}}
\ep^{\mu\nu\rho\sigma}\p_{\rho}
(e^{AA\yp}_{\mu}e_{\nu BA\yp}A_{\sigma A}\,^B+
ie^{AA\yp}_{\mu}
\bar{\psi}_{\nu A\yp}\psi_{\sigma A})
\eeq
So as far as the Euler-Lagrange equations are concerned,
${\cal L}$
and ${\cal L}_J$ are equivalent. But why do we use
${\cal L}$ rather
than ${\cal L}_J$? As discussed in \cite{s13},
conservative quantities
in general relativity are often quasi-local, i.e. can be
expressed as a surface integral at $\p\Sigma$. Whence
total divergences
in the Lagrangian may do non-trivial contribution to the
conservative
quantities though they do not affect the motion
equations. In the
non-supersymmetric case, we also used a different
Lagrangian
in order to obtain the energy-momentum thereof\cite{s7}.
In order to agree with the definition of energy-momentum
in the non-supersymmetric case, in which the energy-momentum
agree exactly with the ADM definition, we use the Lagrangian
(38) here.\\
\indent Since the action $I=\int {\cal L} d^4x$
is invariant under the
infinitesimal transform $x^{\prime\mu}=x^{\mu}+\delta x^{\mu},
\phi^{\prime\ell }_{\mu}(x\yp)=\phi^{\ell}_{\mu}(x)+
\delta\phi^{\ell}_{\mu},
$, here $\phi_{\mu}^{\ell}=e^{AA\yp}_{\mu}, A_{\mu MN},
\bar{\psi}_{\mu A\yp}, \psi_{\mu A}$  we have the
N$\ddot{o}$ether theorem
\beq
\p_{\mu}({\cal L}dx^{\mu}+\delta_0\phi^{\ell}_{\lambda}
\frac{\p{\cal L}}{\p\p_{\mu}\phi^{\ell}_{\lambda}})+
\delta_0\phi^{\ell}
_{\lambda}[{\cal L}]_{\phi^{\ell}_{\lambda}}=0
\eeq
where $ [{\cal L}]_{\phi^{\ell}_{\lambda}}
=(\frac{\p}{\p\phi^{\ell}_{\lambda}}-
\p_{\mu}\frac{\p}{\p\p_{\mu}\phi^{\ell}
_{\lambda}}){\cal L}$ and $\delta_0
\phi^{\ell}_{\lambda}=\delta\phi^{\ell}_{\lambda}
-\p_{\mu}\phi^{\ell}_{\lambda} d^{\mu}x$. Using
the field equations, we
have
\beq
\p_{\mu}({\cal L}dx^{\mu}+\delta_0\phi^{\ell}_{\lambda}
\frac{\p{\cal L}}{\p\p_{\mu}\phi^{\ell}_{\lambda}})=0
\eeq
Since all the fields $\phi^{\ell}_{\mu}$ have a lower
curved index, we have
$\delta\phi^{\ell}_{\nu}=-\delta x^{\mu}_{,\nu}
\phi^{\ell}_{\mu}$. (The
"," denotes "partial derivative"). Therefore,
$\delta_0\phi^{\ell}_{\nu}
=-\delta x^{\lambda}_{,\nu}\phi^{\ell}_{\lambda}-
\p_{\lambda}\phi^{\ell}_{\nu}\delta x^{\lambda}$. Hence we have
\beq
\p_{\mu}[({\cal L}\delta^{\mu}_{\lambda}
-\p_{\lambda}\phi^{\ell}_{\nu}
\frac{\p{\cal L}}{\p\p_{\mu}\phi^{\ell}_{\nu}})
\delta x^{\lambda}
-(\phi^{\ell}_{\lambda}\frac{\p{\cal L}}{\p\p_{\mu}
\phi^{\ell}_{\nu}})\delta
x^{\lambda}_{,\nu}]=0
\eeq
which can be expressed as
\beq
\p_{\mu}[\tilde{I}^{\mu}_{\lambda}\delta x^{\lambda}
+\tilde{V}^{\mu\nu}
_{\lambda}\delta x^{\lambda}_{,\nu}]=0
\eeq
Therefore the independence of $\delta x^{\mu},
\delta x^{\mu}_{,\nu}
$ and $\delta x^{\mu}_{\nu\lambda}$ implies
\beq
\p_{\mu}\tilde{I}^{\mu}_{\lambda}=0,\,\,\,\,\,\,\,\,\,
\tilde{I}^{\nu}_{\lambda}=
-\p_{\mu}\tilde{V}^{\mu\nu}_{\lambda},
\,\,\,\,\,\,\,\,\,\,
\tilde{V}^{\mu\nu}_{\lambda}=
-\tilde{V}^{\nu\mu}_{\lambda}
\eeq
Finally, we use the general translation:
$\delta x^{\mu}=e^{\mu}_{AA\yp}
b^{AA\yp}, b^{AA\yp}$ is an arbitrary infinitesimal
four-vector. This step is cruicial. Note that
 as discussed in \cite{s7}, the general coordinate transform
 $\delta x^{\mu}=\xi^{\mu}$ does not
 in fact effect a translation because $x^{\mu}$ can be any
 curvilinear coordinate. Now we have
\beq
\p_{\mu}(\tilde{I}^{\mu}_{\lambda}
e^{\lambda}_{AA\yp}+\tilde{V}
^{\mu\nu}_{\lambda}\p_{\nu}e^{\lambda}_{AA\yp})=0
\eeq
The energy-momentum tensor is defined to be
\beq
\tilde{t}^{\mu}_{AA\yp}:=et^{\mu}_{AA\yp}:=
\tilde{I}^{\mu}_{\lambda}e^{\lambda}_{AA\yp}+\tilde{V}
^{\mu\nu}_{\lambda}\p_{\nu}e^{\lambda}_{AA\yp}
\eeq
So
\beq
\p_{\mu}(et^{\mu}_{AA\yp})=0
\eeq
Using $\tilde{I}^{\mu}_{\lambda}
=\p_{\nu}\tilde{V}^{\mu\nu}_{\lambda}$, we
have
\beq
\tilde{t}^{\mu}_{AA\yp}=\p_{\nu}\tilde{V}^{\mu\nu}_{AA\yp},\,
\,\,\,\,\,\,
\tilde{V}^{\mu\nu}_{AA\yp}:=\tilde{V}^{\mu\nu}_{\lambda}
e^{\lambda}_{AA\yp}
\eeq
Since
\beq
\frac{\p{\cal L}}{\p\p_{\mu}e^{AA\yp}_{\nu}}
=-\sqrt{2}\ep^{\mu\nu\alpha
\beta}e_{\alpha BA\yp}A_{\beta A}\,^B
-\frac{i}{2\sqrt{2}}\ep
^{\mu\nu\alpha\beta}\bar{\psi}_{\alpha A\yp}\psi_{\beta A}
\eeq
\beq
\frac{\p{\cal L}}{\p\p_{\mu}A_{\nu MN}}=0,\,\,\,\,\,\,\,
\frac{\p{\cal L}}{\p\p_{\mu}\bar{\psi}_{\nu A\yp}}
=\frac{i}{2\sqrt{2}}
\ep^{\mu\nu\alpha\beta}e_{\alpha}\,^{AA\yp}\psi_{\beta A}
\eeq
\beq
\frac{\p{\cal L}}{\p\p_{\mu}\psi_{\nu A}}
=-\frac{i}{2\sqrt{2}}
\ep^{\mu\nu\alpha\beta}e_{\alpha}\,^{AA\yp}
\bar{\psi}_{\beta A\yp}
\eeq
we have
\beq
\tilde{V}^{\mu\nu}_{\lambda}=
\ep^{\mu\nu\alpha\beta}[\sqrt{2}
e_{\lambda}\,^{AA\yp}e_{\alpha BA\yp}A_{\beta A}\,^B
+\frac{i}{2\sqrt{2}}
(e_{\lambda}^{AA\yp}\bar{\psi}_{\alpha A\yp}\psi_{\beta A}-
e_{\alpha}\,^{AA\yp}\bar{\psi}_{\lambda A\yp}\psi_{\beta A}-
e_{\alpha}\,^{AA\yp}\bar{\psi}_{\beta A\yp}\psi_{\lambda A})]
\eeq
and
$$
\tilde{V}^{\mu\nu}_{NN\yp}=\ep^{\mu\nu\alpha\beta}[\sqrt{2}
e_{\lambda}\,^{AA\yp}e_{\alpha BA\yp}e^{\lambda}_{NN\yp}
A_{\beta A}\,^B+\frac{i}{2\sqrt{2}}
(e_{\lambda}^{AA\yp}e^{\lambda}_{NN\yp}
\bar{\psi}_{\alpha A\yp}\psi_{\beta A}$$
\beq
-e_{\alpha}\,^{AA\yp}\bar{\psi}_{\lambda A\yp}\psi_{\beta A}
e^{\lambda}_{NN\yp}
-e_{\alpha}\,^{AA\yp}\bar{\psi}_{\beta A\yp}\psi_{\lambda A}
e^{\lambda}_{NN\yp})]
\eeq
For a closed system, the conservative energy-momentum is
\beq
P_{NN\yp}=\int_{\Sigma} et^0_{NN\yp} d^3x=\int_{\Sigma}
\p_i\tilde{V}^{0i}_{NN\yp} d^3x=\int_{\p\Sigma}
\tilde{V}^{0i}_{NN\yp}ds_i
\eeq
where
$$
\tilde{V}^{0i}_{NN\yp}
=\ep^{ijk}[\sqrt{2}
e_{\lambda}\,^{AA\yp}e_{jBA\yp}e^{\lambda}_{NN\yp}
A_{kA}\,^B+\frac{i}{2\sqrt{2}}
(e_{\lambda}^{AA\yp}e^{\lambda}_{NN\yp}
\bar{\psi}_{jA\yp}\psi_{kA}$$
\beq
-e_{j}\,^{AA\yp}\bar{\psi}_{\lambda A\yp}\psi_{kA}
e^{\lambda}_{NN\yp}
-e_{j}\,^{AA\yp}\bar{\psi}_{kA\yp}\psi_{\lambda A}
e^{\lambda}_{NN\yp})]
\eeq
Now we use the reality of $P_{AA\yp}$ to simplify the expression.
Since both sides of eq.(4) are real, the first term is real.
It is also because of the reality of
$i\psi_A\wedge\bar{\psi}_{A\yp}
$, the last two terms contribute nothing to the
real $P_{AA\yp}$.
So we have
\beq
P_{NN\yp}=\int_{\p\Sigma}\ep^{ijk}[\sqrt{2}
e_{\lambda}\,^{AA\yp}e_{jBA\yp}e^{\lambda}_{NN\yp}
A_{kA}\,^B+\frac{i}{2\sqrt{2}}
e_{\lambda}^{AA\yp}e^{\lambda}_{NN\yp}
\bar{\psi}_{jA\yp}\psi_{kA}]ds_i
\eeq
From eq.(4) we have $\bar{\psi}_{[jA\yp}\psi_{k]A}
=2i({\cal D}_{[j}e_{k]})_{AA\yp}$. therefore
\beq
P_{NN\yp}=\frac{1}{\sqrt{2}}\int_{\p\Sigma}\ep^{ijk}
e_{\lambda}\,^{AA\yp}e_{jBA\yp}e^{\lambda}_{NN\yp}
A_{kA}\,^B ds_i
\eeq
\indent To make the 3+1 decomposition of $P_{AA\yp}$,
we may use two ways. The first one is to use
 the relationship between the flat $SL(2,\Bbb C)$ soldering
form $E^a_{AA\yp}$ and the sigma matrices in ref.\cite{s14}
\beq
E^a_{AA\yp}=\frac{1}{\sqrt{2}}\sigma^a_{AA\yp}
\eeq
and
\beq
\sigma^{aAA\yp}\sigma_{a BB\yp}=2\delta^A_B\delta^{A\yp}_{B\yp}
\eeq
we have
\beq
e^{AA\yp}_{\lambda}e^{\lambda}_{NN\yp}=
e^a_{\lambda}e^{\lambda}_b E_a^{AA\yp}E^b_{NN\yp}=
\delta^A_N\delta^{A\yp}
_{N\yp}
\eeq
so
\beq
P_{AA\yp}=\frac{1}{\sqrt{2}}
\int_{\p\Sigma}\ep^{ijk}e_{jBA\yp}A_{kA}\,^B ds_i
\eeq
Use $P_a=E_a^{AA\yp}P_{AA\yp}$ and
\beq
E_a^{AA\yp}=-i\sqrt{2}E_a^{AB}n_B\,^{A\yp}+n_an^{AA\yp}
\eeq
we have
\beq
P_0=P_{NN\yp}n^{NN\yp}=\frac{1}{\sqrt{2}}
\int_{\p\Sigma}\ep^{ijk}e_{jBA\yp}A_{kA}\,^B n^{AA\yp}ds_i
=\frac{-i}{2}\int_{\p\Sigma}\ep^{ijk}e_{jB}
\,^NA_{kN}\,^B ds_i
\eeq
i.e.
\beq
P_0=\frac{-i}{2}\int_{\p\Sigma}\ep^{ijk}\tr e_jA_k ds_i
\eeq
\beq
P_m=P_{NN\yp}(-i\sqrt{2})E_m\,^{NC}n_C\,^{N\yp}
=\frac{1}{\sqrt{2}}\int_{\p\Sigma}\ep^{ijk}(A_je_k)_{MN}
E_m^{MN}ds_i
=\frac{1}{2}E_m^{MN}P_{MN}
\eeq
The second way is to use
\beq
e_{\lambda}\,^{AA\yp}e^{\lambda}_{NN\yp}=2e_{\lambda}\,^{AB}
e^{\lambda}\,_N\,^Mn_B\,^{A\yp}n_{MN\yp}+n^{AA\yp}n_{NN\yp}
\eeq
and
\beq
e^0_{MN}=0,\,\,\,\,\,\, e^i_{AB}e_i^{MN}
=\delta_A^{(M}\delta_B^{N)}
\eeq
(Note that $e^{\mu}_{AA\yp}:=g^{\mu\nu}e_{\nu AA\yp}$ while
$e^{\mu}_{AB}:=-g^{\mu\nu}e_{\nu AB}, n_{\mu}=(n_0,0,0,0)$).
Substituting the first term which is spatial of the
r.h.s of eq.(66)
into eq(57) gives the momentum $P_m$ and  the second term
gives the energy $P_0$. The result is the same.\\
\indent We now rescale the energy-momentum by
a constant factor $2\sqrt{2}$.
i.e.
\beq
P_{AA\yp}=2\int_{\p\Sigma}\ep^{ijk}e_{jBA\yp}A_{kA}\,^B ds_i
\eeq
\beq
P_0=-i\sqrt{2}\int_{\p\Sigma}\ep^{ijk}\tr e_jA_k ds_i
\eeq
\beq
P_m=2\int_{\p\Sigma}\ep^{ijk}(A_je_k)_{MN}E_m^{MN}ds_i
=\frac{1}{\sqrt{2}}E_m^{MN}P_{MN}
\eeq
so that it agrees exactly with that of the non-sypersymmetric
case\cite{s7}. \\
\indent The Poisson bracket $\{J_{MN},P_{AB}\}$ can be
calculated
using $(e^ie_i)_{(MN)}=0, \sqrt{2}e^{[j}e^{k]}=
q^{-1/2}\ep^{ijk}e_i$
and
\beq
\{e^{AB}_i(x),A_{jMN}(y)\}=
2q^{-1/2}e^{AB}_{[i}e_{j]MN}\delta^3(x,y)
-\frac{1}{2\sqrt{2}}q^{-1}
\ep_{lmj}(\ts^l\ts^m)_{(MN)}e_i^{AB}\delta^3(x,y)
\eeq
The result is
\beq
\{J_{MN},P_{AB}\}=\frac{1}{2}(\ep_{AM}P_{NB}+\ep_{BM}P_{AN}
+\ep_{AN}P_{MB}+\ep_{BN}P_{MA})
\eeq
therefore
\beq
\{J_m,P_n\}=\{\frac{1}{\sqrt{2}}E_m^{AB}J_{AB},
\frac{1}{\sqrt{2}}
E_n^{MN}P_{MN}\}=\ep_{mnl} P^l
\eeq
\indent To calculate $\{P_{MN},P_{AB}\}$, one of
the two $P$'s must be
expressed as a 3-dim integral.
\beq
P_{AB}=\int_{\p\Sigma}4q^{-1/2}(A_j\ts^{[i}
\ts^{j]})_{(AB)}ds_i
=\int_{\Sigma}4\p_i(q^{-1/2}(A_j\ts^{[i}
\ts^{j]})_{(AB)}) d^3x
\eeq
As in the non-supersymmetric case\cite{s7},
it is not differentiable
with respect to $A_{iMN}$. To circulmvent this difficulty,
we use the same trick as in \cite{s7}, which stems from
the construction of the Hamiltonian generating time
translations\cite{s1}-\cite{s2}.
Suppose $\underline{N}$ is a scalar density of weight $-1$
and equals
$q^{-1/2}$ outside a compact set of $\Sigma$. So
\beq
\begin{array}{rcl}
P_{AB}&=&\int_{\Sigma}4
          [\p_i(\underline{N}A_j\ts^{[i}
          \ts^{j]})]_{(AB)}d^3x\\\\
&=&\int_{\Sigma}4[\p_i(\underline{N}\ts^{[i}\ts^{j]})A_j
+\underline{N}\ts^{[i}\ts^{j]}(\frac{1}{2}F_{ij}-
A_iA_j)]_{(AB)}d^3x
\end{array}
\eeq
Use the constraint eq(15), we have
\beq
\begin{array}{rcl}
P_{AB}&\approx&\int_{\Sigma}4[\p_i(\underline{N}
\ts^{[i}\ts^{j]})A_j
+\underline{N}\ts^{[i}\ts^{j]}(\frac{1}{2}F_{ij}-A_iA_j)
-\frac{1}{2}\underline{N}\ddot{\cal H}]_{(AB)}d^3x\\\\
&=&\int_{\Sigma}4[\p_i(\underline{N}\ts^{[i}\ts^{j]})A_j-
\underline{N}\ts^{[i}\ts^{j]}A_iA_j-\underline{N}
\tp^j\ts^k{\cal D}_{[j}\psi_{k]}\ep-\underline{N}
\tp^j{\cal D}_{[j}\psi_{k]}\ts^k]_{(AB)}d^3x\\\\
&=&\int_{\Sigma}4[\p_i(\underline{N}\ts^{[i}\ts^{j]})A_j-
\underline{N}\ts^{[i}\ts^{j]}A_iA_j
-\underline{N}\tp^j{\cal D}_{[j}\psi_{k]}\ts^k]_{(AB)}d^3x
\end{array}
\eeq
When taking into consideration the falloff $d\ts\sim
r^{-2}, A\sim r^{-2},
\psi,\bar{\psi},\tp \sim r^{-1}$, we have
\beq
\{P_{AB},P_{MN}\}\approx 0
\eeq
i.e.
\beq
\{P_i,P_j\}\approx 0
\eeq
\begin{center}
\section{The Supercharges}
\end{center}
Since the Lagrangian varies as
\beq
\delta{\cal L}=\delta\phi^{\ell}_{\mu}
[{\cal L}]_{\phi^{\ell}_{\mu}}
+\p_{\mu}(\delta\phi^{\ell}_{\nu}
\frac{\p{\cal L}}{\p\p_{\mu}\phi^{\ell}
_{\nu}})
\eeq
we have on-shell that
\beq
\delta{\cal L}=\p_{\mu}(\delta\phi^{\ell}_{\nu}
\frac{\p{\cal L}}{\p\p_{\mu}\phi^{\ell}_{\nu}})
\eeq
On the other hand, one can calculate $\delta{\cal L}$
directly.Using
eq(39) and the invariance of ${\cal L}_J$ under the
transform eq.(5),
we have the variation of $\cL$ under the left-handed
transform eq.(5).
\beq
\delta\cL=\sqrt{2}i\ep^{\mu\nu\rho\sigma}\p_{\rho}
(e^{AA\yp}_{\mu}\bar{\psi}_{\nu A\yp}A_{\sigma A}\,^B\ep_B-
\bar{\psi}_{\nu A\yp}e^{AA\yp}_{\mu}{\cal D}_{\sigma}\ep_A)
\eeq
Whence
\beq
\p_{\mu}(\delta\phi^{\ell}_{\nu}
\frac{\p{\cal L}}{\p\p_{\mu}\phi^{\ell}_{\nu}})
=-\sqrt{2}i\ep^{\mu\nu\rho\sigma}\p_{\rho}\
(e^{AA\yp}_{\mu}\bar{\psi}_{\nu A\yp}\p_{\sigma}\ep_A)
\eeq
Since
\beq
\begin{array}{rcl}
\p_{\mu}(\delta\phi^{\ell}_{\nu}
\frac{\p{\cal L}}{\p\p_{\mu}\phi^{\ell}_{\nu}})
&=&\ep^{\alpha\beta\mu\nu}\p_{\mu}[(-\sqrt{2}
e_{\alpha BA\yp}
A_{\beta A}\,^B-\frac{i}{2\sqrt{2}}
\bar{\psi}_{\alpha A\yp}\psi_{\beta A})
(-i\bar{\psi}_{\nu}^{A\yp}\ep^A)\\\\
& &+i\frac{1}{\sqrt{2}}e_{\alpha}\,^{AA\yp}
\bar{\psi}_{\beta A\yp}{\cal D}_{\nu}\ep_A]\\\\
&=&\ep^{\alpha\beta\mu\nu}\p_{\mu}[i\sqrt{2}
e_{\alpha BA\yp}
A_{\beta A}\,^B\bar{\psi}^{A\yp}_{\nu}\ep^A
+i\frac{1}{\sqrt{2}}e_{\alpha}\,^{AA\yp}
\bar{\psi}_{\beta A\yp}{\cal D}_{\nu}\ep_A]
\end{array}
\eeq
we have
\beq
\ep^{\alpha\beta\mu\nu}\p_{\mu}[i\sqrt{2}
e_{\alpha BA\yp}
A_{\beta A}\,^B\bar{\psi}^{A\yp}_{\nu}\ep^A
+i\frac{1}{\sqrt{2}}e_{\alpha}\,^{AA\yp}
\bar{\psi}_{\beta A\yp}{\cal D}_{\nu}\ep_A
+\sqrt{2}ie_{\alpha}\,^{AA\yp}
\bar{\psi}_{\beta A\yp}\p_{\nu}\ep_A]=0
\eeq
Therefore. we have
\beq
\p_{\mu}(\tilde{Q}^{\mu}_B\ep^B+
\tilde{Q}^{\mu\nu}_B\p_{\nu}\ep^B)=0
\eeq
where
\beq
\tilde{Q}^{\mu}_B:=2\ep^{\mu\nu\alpha\beta}
e_{\alpha AA\yp}A_{\beta B}
\,^A\bar{\psi}^{A\yp}_{\nu},\,\,\,\,\,\,\,\,
\tilde{Q}^{\mu\nu}_B:=-2\ep^{\mu\nu\alpha\beta}
e_{\alpha B}\,^{B\yp}
\bar{\psi}_{\beta B\yp}
\eeq
The independence of $\ep_A, \p_{\mu}\ep_A$ and
$\p_{\mu\nu}\ep_A$
implies that
\beq
\p_{\mu}\tilde{Q}^{\mu}_A=0,\,\,\,\,\,\,\,\,
\tilde{Q}^{\nu}_B=\p_{\mu}\tilde{Q}^{\nu\mu}_B,
\,\,\,\,\,\,\,\,
\tilde{Q}^{\mu\nu}_B=-\tilde{Q}^{\nu\mu}_B
\eeq
So we have the left-handed supercharge
\beq
Q_A=2\int_{\Sigma}\tilde{Q}^0_A d^3x=
2\int_{\p\Sigma}\tilde{Q}
^{0i}_A ds_i=2\int_{\p\Sigma}\ep^{ijk}e_{jAA\yp}
\bar{\psi}_{k}^{A\yp} ds_i
=i2\sqrt{2}\int_{\p\Sigma}\tp^i_A ds_i
\eeq
\indent To obtain the right-handed supercharge,
we use the right-handed
transform eq.(6) under which $\cL_J$ transforms as
\beq
\cL_J=-{\cal D}(\sqrt{2}ie^{AA\yp}\wedge
\bar{\ep}_{A\yp}{\cal D}\psi_A)
+(\sqrt{2}i{\cal D}e^{AA\yp}+\psi^A\wedge
\bar{\psi}^{A\yp})
\wedge\bar{\ep}_{A\yp}{\cal D}\psi_A
\eeq
Using the field equation $e^{AA\yp}\wedge{\cal D}
\psi_A=0$ and
eq.(4), we have $\delta\cL_J=0$. Thus
\beq
\begin{array}{rcl}
\delta\cL&=&-\ep^{\mu\nu\rho\sigma}\p_{\rho}
[-\sqrt{2}i\psi^A_{\mu}
\bar{\ep}^{A\yp}e_{\nu BA\yp}A_{\sigma A}\,^B+
\frac{1}{\sqrt{2}}
\psi^A_{\mu}\bar{\ep}^{A\yp}\bar{\psi}_{\nu A\yp}
\psi_{\sigma A}\\\\
& &+\sqrt{2}ie_{\mu}^{AA\yp}{\cal D}_{\nu}
\bar{\ep}_{A\yp}\psi_{\sigma A}]\\\\
&=&-\ep^{\mu\nu\rho\sigma}\p_{\rho}[-\sqrt{2}i\psi^A_{\mu}
\bar{\ep}^{A\yp}e_{\nu BA\yp}A_{\sigma A}\,^B+
\sqrt{2}ie_{\mu}^{AA\yp}{\cal D}_{\nu}
\bar{\ep}_{A\yp}\psi_{\sigma A}]=0
\end{array}
\eeq
This can yield that
\beq
\p_{\mu}(\tilde{\bar{Q}}^{\mu}_{A\yp}\bar{\ep}^{A\yp}
+\tilde{\bar{Q}}^{\mu\nu}_{A\yp}\p_{\nu}\bar{\ep}^{A\yp})
=0
\eeq
where
\beq
\tilde{\bar{Q}}^{\rho}_{A\yp}:=-2\ep^{\mu\nu\rho\sigma}
\psi^A_{\mu}e_{\nu BA\yp}
A_{\sigma A}\,^B,\,\,\,\,\,\,\,
\tilde{\bar{Q}}^{\rho\nu}_{A\yp}:=-2\ep^{\mu\nu\rho\sigma}
e_{\mu AA\yp}
\psi^A_{\sigma}
\eeq
   Similar to eq.(87), we have
\beq
\p_{\mu}\tilde{\bar{Q}}^{\mu}_{A\yp}=0,\,\,\,\,\,\,\,\,
\tilde{\bar{Q}}^{\nu}_{B\yp}=\p_{\mu}
\tilde{\bar{Q}}^{\nu\mu}_{B\yp},\,\,\,\,\,\,\,\,
\tilde{Q}^{\mu\nu}_{B\yp}
=-\tilde{\bar{Q}}^{\nu\mu}_{B\yp}
\eeq
and the right-handed supercharge is
\beq
{\bar{Q}}_{A\yp}=\int_{\Sigma}
\tilde{\bar{Q}}^0_{A\yp} d^3x=\int_{\p\Sigma}
\tilde{\bar{Q}}^{0i}_{A\yp}ds_i=
-2\int_{\Sigma}\ep^{ijk}
\psi^A_ie_{j BA\yp}A_{kA}\,^B d^3x=2
\int_{\p\Sigma}\ep^{ijk}e_{jAA\yp}\psi^A_k ds_i
\eeq
We can easily see that $Q_A$ and ${\bar{Q}}_{A\yp}$
are complex
conjugate to each other by comparing eq.(88)
and eq.(94). Their
Poisson bracket gives that
\beq
\{Q_A,\bar{Q}_{A\yp}\}
=\{i2\sqrt{2}\int_{\p\Sigma}\tp^l_Ads_l,
-2\int_{\Sigma}\ep^{ijk}\psi^C_ie_{jBA\yp}A_{kC}\,^Bd^3x\}
=i2\sqrt{2}P_{AA\yp}=i2\sigma_{AA\yp}^aP_a
\eeq
\indent Using the volume integral of the supercharges and
the surface
integral of the energy-momentum and taking into the
fall-off of the fields, one may easily obtain that
\beq
\{Q_A,\,\,\,P_{BB\yp}\}=\{\bar{Q}_{A\yp},\,\,\,
P_{BB\yp}\}=0
\eeq
\indent Finally, we calculate the Poisson bracket
of the supercharges
and the $SU(2)$ charges. Note that the quantal
commutator of them is
\cite{s15}
\beq
[Q_A,\,\,\, J_{ab}]_-=\frac{1}{2}(\sigma_{ab})_A\,^BQ_B
\eeq
where $(\sigma_{ab})_A\,^B=-\frac{1}{2}(\sigma_{aA}
\,^{B\yp}\sigma_b\,^B\,
_{B\yp}-\sigma_{bA}\,^{B\yp}\sigma_a\,^B\,_{B\yp})$
(here the
$\sigma$-matrices are those in\cite{s14} not in
\cite{s15} in which
the $\sigma$-matrices with one lower primed index
differ  by a sign
from those in \cite{s14}). So
\beq
[Q_A,\,\,\, J_{ab}]_-=-E_{[aA}\,^{B\yp}E_{b]}\,^B\,_{B\yp}
\eeq
Using
\beq
E_{[aA}\,^{B\yp}E_{b]}\,^B\,_{B\yp}=E_{[aAC}E_{b]}
\,^{BC}-i\sqrt{2}
n_{[a}E_{b]A}\,^B
\eeq
we have
\beq
[Q_A,\,\,\, J_{ij}]=-E_{[iAC}E_{j]}\,^{BC}Q_B
\eeq
and
\beq
[Q_A,\,\, J_{0i}]_-=i\sqrt{2}n_{[0}E_{i]A}\,^BQ_B
=\frac{i}{\sqrt{2}}
E_{iA}\,^BQ_B
\eeq
On the other hand, we have from eq(32) that
\beq
\tilde{j}^{\rho}_{AB}=-\frac{1}{2}\tilde{j}^{\rho}_{ab}
E^a\,_A\,^{A\yp}E^b\,_{BA\yp}
\eeq
where $\tilde{j}^{\rho}_{ab}$ is the angular-momentum current
obtained in \cite{s9}-\cite{s10}.
\beq
\tilde{j}^{\rho}_{ab}=\sqrt{2}\ep^{\rho\sigma\mu\nu}\p_{\sigma}
(e_{\mu a}e_{\nu b})
\eeq
and the angualr-momentum is
\beq
J_{ab}=\int_{\Sigma}\tilde{j}^0_{ab} d^3x
\eeq
Hence
\beq
\begin{array}{rcl}
J_{MN}&=&-\frac{1}{2}J^{ab}E_{[aM}\,^{A\yp}E_{b]NA\yp}=
-\frac{1}{2}(J^{ij}E_{[iAC}E_{j]}\,^{BC}-
i\sqrt{2}J^{0i}n_0E_{iA}\,^B)\\\\
&=&\frac{1}{\sqrt{2}}(L_i-iK_i)E^i_{MN}
\end{array}
\eeq
where $L_i=\frac{1}{2}\ep_{ijk}J^{jk}$ are the
spatial rotations
and $K_i=J_{0i}=-J^{0i}$ are the Lorentz boosts.
Therefore
\beq
J_i=\frac{1}{2}(L_i-iK_i)
\eeq
(Bearing in mind that both $\frac{1}{2}(L_i-iK_i)$
and $\frac{1}{2}(L_i+iK_i)$ obey the $su(2)$
algbra\cite{s16}-\cite{s17})
From eq.(100) and eq. (101) we have
\beq
[Q_A,\,\,\, J_k]_-=\frac{1}{\sqrt{2}}E_{kA}\,^BQ_B
\eeq
This can really be realized by the Poisson bracket because
\beq
\{Q_A,\,\, J_i\}=\{Q_A,\,\,\, \frac{1}{\sqrt{2}}
E_i^{MN}\int_{\p\Sigma}\ts^k_{MN} ds_k\}
=\frac{1}{\sqrt{2}}E_{iA}\,^BQ_B
\eeq
Actually, the boost charges are vanishing here as can be
seen from eq(30).
\begin{center}
\section{Summary and Discussions}
\end{center}
In this paper, we have obtained the angualr-momentum,
supercharges and
the energy-momentum in the self-dual simple supergravity.
The conservation laws possess the common feature of
the conservation laws
obtained previously, i.e., the  currents are identically
conservative because they can expressed as  divergences
of  antisymmetric tensor densities which are
often referred to
as {\it potentials}.
The total charges take the same integral forms
as those in the
non-supersymmetric case. Though we can obtain
the $SU(2)$
sector of the $SL(2,\Bbb C)$ charges, the information
of the angular-momentum is completely contained in the
$SU(2)$ charges. It can be seen from the surface
integrals
that the angular-momentum
is governed by the $r^{-2}$ part of $\ts^i$, the
energy-momentum
is determined by the $r^0$ part of $\ts^i$ and
the $r^{-2}$ part
of $A_i$, and the supercharges are governed by
the $r^{-2}$ part
of $\tp^i$. As in \cite{s1}-\cite{s2}, we always
assume that
the phase space variables are subject to the boundary
conditions.
\beq
e^{\mu}_{AB\mid\p\Sigma}=
(1+\frac{M(\theta,\phi)}{r})^2\stackrel{0}{e
^{\mu}}_{AB}+O(1/r^2),\,\,\,\,
A_{\mu MN}\,_{\mid\p\Sigma}=O(1/r^2)
\eeq
\beq
\tp^i_A=O(1/r),\,\,\,\, \psi_{\mu A}=O(1/r)
\eeq
where $\stackrel{0}{e^{\mu}}_{AB}$ denote the
flat $SU(2)$ soldering forms.
As a consequence, under the $SL(2,\Bbb C)$
transforms behaving as
\beq
L_A\,^B=\Lambda_A\,^B+O(1/r^{1+\ep}), \,\,\,\,\,\,(\ep>0)
\eeq
where $\Lambda$ are rigid transforms
The charges transform as
\beq
J_{MN}\rightarrow\Lambda_M\,^A\Lambda_N\,^BJ_{AB},\,\,\,
P_{AA\yp}\rightarrow\Lambda_A\,^{B}\bar{\Lambda}
_{A\yp}\,^{B\yp}P_{BB\yp}
\eeq
\beq
Q_A\rightarrow \Lambda_A\,^BQ_B,\,\,\,\,\,\,
\bar{Q}^{A\yp}\rightarrow \bar{\Lambda}_{A\yp}\,^{B\yp}
Q_{B\yp}
\eeq
i.e., they gauge covariant.
Their conservation is generally covariant. Upon quantization,
the Poisson brackets correspond to the quantal commutators
or anti-commutators\cite{s18}-\cite{s19}  and their algebra
realizes indeed the super-Poincare algebra. This shows that
their interpertations convincing, especially that the
approaches used previously to obtain generally covariant
conservation  laws are reasonable.\\
\indent It is novel that the relationship among
the conservative
quantities and the first class constraints is the
same as the
gauge charges and the constraints in the usual
Yang-Mills gauge field models. To see this,
consider the example of interacting Yang-Mills
and spinor fields\cite{s20}. The Lagrangian
\beq
\cL=-\frac{1}{4}{\cal F}_{\mu\nu}^a{\cal F}^{\mu\nu}_a
+i\bar{\psi}\gamma^{\mu}D_{\mu}\psi-m\bar{\psi}\psi
\eeq
is invariant under gauge transforms and this leads
to the conservative Noether currents
\beq
J^{\mu}_a=t^{abc}{\cal F}^{\mu\nu}_b
{\cal A}_{\nu}^c+i\bar{\psi}\gamma^{\mu}T^c\psi
\eeq
where $[T^a,T^b]=t^{abc}T^c$. Among the equations
of motion, there are the constraints
\beq
C^a(x)=\p^k{\cal F}^a_{k0}-t^{abc}{\cal A}_b^k{\cal F}
^c_{0k}+i\bar{\psi}\gamma_0T^a\psi\approx 0
\eeq
 generating time-independent
 gauge transforms.
 The zero component of $J^{\mu}_a$ is just the
 the last two terms of the constraints.
 So we have the gauge charges
\beq
Q_a=\int_{\Sigma} J^0_a d^3x=\int_{\p\Sigma} {\cal F}_a^{0k} ds_k
\eeq
i.e., it is also a surface integral modulo the constraints.
The surface integral expressions of $J_{MN}, Q_A$ et al
in this paper
are also obtained in this way.\\
\indent The supergravity considered here is an extension
of the
non-supersymmetric case. This can be seen by setting
the anticommuting
fields to be zero. Then not only the field equations but also
the constraints reduces to the constraints in \cite{s1}-\cite{s2}.
The Hamiltonian constriant and the diffeomorphism constraint
in \cite{s1}-\cite{s2} are implied in the constraint $ {\cal H}
^{AA\yp}$. Since eq(15) which stems from eq(11)
reduces to the Hamiltonian
constraint
\beq
\tr(\ts^i\ts^jF_{ij})=0
\eeq
and eq(11) reduces to
\beq
\ep^{ijk}e_i\,^{BD}F_{jkD}\,^A=0
\eeq
so
\beq
\ep^{ijk}e_i\,^{BD}F_{jkD}\,^AE_{lAB}=0
\eeq
i.e.
\beq
\ep^{ijk}e_i^m(E_mE_l)_A\,^DF_{jkD}\,^A=0
\eeq
Using $(E_mE_l)_A\,^D=\frac{1}{2}(\delta_{ml}\delta_A\,^D+
\ep_{mln}E_{nA}\,^D), \tr F_{ij}=0 $ and
$\ep^{ijk}\ep_{mln}e^m_i\sim
e^{[j}_le^{k]}_n$,  we can obtain the diffeomorphism
constraint
\beq
\tr (\ts^iF_{ij})=0
\eeq
Thus we can say that the Hamiltonian constraint
and the diffeomorphism
constraint can be combined together.
\vskip 0.3in

\underline{\bf Acknowledgement} S.S. Feng is indebted to Prof. S.
Randjbar-Daemi for his invitation for three months at ICTP.
This work is
supported by the National Sceince
Foundation of China under Grant. No. 19805004 and in part by the Funds for
Young Teachers of Shanghai Education Council.


\begin{thebibliography}{s40}
\bibitem{s1} A Ashtekar {\it Phys. Rev. Lett.} {\bf 57}
            (1986):2244; {\it Phys. Rev. }{\bf D 36}
            (1987):1587.
\bibitem{s2} A.Ashtekar {\it New Perspectives in
           Canonical Gravity}
            (Lecture Notes, 1988, Naples:Biblipolis).
\bibitem{s3} T. Jacobson \& L.Smolin {\it Phys. Lett}
             {\bf B 196}(1987):39.
\bibitem{s4} J. Samuel {\it Pramana J Phys.} {\bf 28}
          (1987):L429.
\bibitem{s5} T. Jacobson {\it Class. Quan. Grav.}{\bf 5}
            (1988):923.
\bibitem{s6} N.N. Gorobey \& A.S. Lukyanenko {\it Class.
            Quan. Grav.}
            {\bf 7} (1990):67.
\bibitem{s7} S.S. Feng \& Y.S. Duan {\it Gen. Rel. Grav.}
            {\bf 27}(8)
            (1995):887.
\bibitem{s8} S.S. Feng \& Y.S. Duan {\it Commu. Theor. Phys.}
            {\bf 25}(1996):485.
\bibitem{s9} Y.S. Duan \& S.S. Feng {\it Commu. Theor. Phys.}
             {\bf 25} (1996):99.
\bibitem{s10}S.S. Feng \& H.S. Zong  {\it Inter. J. Theor.Phys.}
             {\bf 35} (1996):267. S.S. Feng \& Y.S. Duan
             {\it Grav. \& Cos.} {\bf 1} (1995):319.
\bibitem{s11}S.S. Feng {\it Nucl. Phys. }{\bf B 468}(1996):163.
\bibitem{s12} Y.S. Duan \& J.Y. Zhang {\it Acta. Phys. Sini}
            {\bf 19} (1963):589.
\bibitem{s13} S.S. Feng \& X.J. Qiu {\it Phys. Lett.} {\bf B 411}
              (1997):256.
\bibitem{s14} J.Wess \& J. Bagger {\it
              Supersymmetry and Supergravity}
              second edition. (Princeton University Press, 1992).
\bibitem{s15} Peter West {\it Introduction to Supersymmetry and
             Supergravity} (Extended  Second Edition) 1990, World
              Scientific Publishing Co.Pte.Ltd.
\bibitem{s16} L.H. Ryder {\it Quantum Field Theory}
              (Cambridge University Press,1985)
\bibitem{s17} S. Weinberg {\it The Quantum Theory of Fields}
              Vol.I. (Cambridge University Press,1995).
\bibitem{s18} R.Casalbuoni {\it Nuovo Cim,} {\bf 33A} (1976):115;
              {\it Nuovo Com.} {\bf 33A} (1976):389.
\bibitem{s19} Peter. G. . Freund {\it Introduction to
             Supersymmetry} ( Cambridge University Press, 1986).
\bibitem{s20} L.D. Faddeev \& A.A.Slavnov {\it
              Gauge Fields Introduction to Quantum
              Theory} (Addison-Wesley Publishing Company
              ,1991).
\end{thebibliography}
\end{document}